\documentclass[11pt,twoside]{article}

\usepackage{asp2006}
\usepackage{epsf}
\usepackage{lscape}

\begin{document}
\title{Emitting gas regions in Mrk 493: An extensive Fe II line emission region }
\author{L. \v C. Popovi\' c$^1$,  A. Smirnova$^2$, D.
Ili\'c$^3$,  A. Moiseev$^2$,
J. Kova\v cevi\'c$^1$ \& V. Afanasiev$^2$}   
\affil{$^1$Astronomical Observatory, Volgina 7, 11160 Belgrade 74, Serbia
\\
$^2$Special Astrophysical Observatory,   Russian Academy of Sciences, 369167
Nizhnij Arkhyz, Russia
\\
$^3$ Department of Astronomy, Faculty of Mathematics,
University of Belgrade,  Studentski trg 16, 11000  Belgrade,
Serbia}

\begin{abstract} 
We performed 3D spectroscopic observations of Mrk 493 in order to
investigate  the  Fe II emitting region and their possible connection with the  Hydrogen emitting region. We found that
there is a strong Fe II emission in an extensive region $\sim
4''\times4''$ around Sy 1 nucleus. The Fe II line width indicates
that these lines are originated in an intermediate line region.
\end{abstract}


\section{Introduction}

 Mrk 493, a narrow-line Seyfert 1 galaxy (NLS1), has a strong Fe II
emission. It was shown by  Boroson \& Green (1992) that  a strong
anticorrelation between the strengths of the ${\rm [OIII]}$ and Fe
II exists in the optical  spectra, with NLS1 as a class showing
the strongest Fe II and weakest ${\rm [O III]}$ emission. The
observed line widths and absence of forbidden emission suggests
that Fe II is formed in the dense broad-line region (BLR), but
photoionization models cannot account for all of the Fe II
emission. The 'Fe II discrepancy' remains unsolved, though models
which consider non-radiative heating, with an overabundance of
iron are promising (see paper of Joly in this proceedings). We
performed 3D spectroscopic observations of Mrk 493 in order (i) to
find the structure of the Fe~II  emitting region and possible
connection with other emitting regions and (ii) to map the
circum-nuclear emitting gas.

\section{Observations}

Mrk 493 was observe in August 2002 and May 2004  with the
integral-field  MultiPupil Fiber Spectrograph  (MPFS), mounted at
the primary focus  of the SAO RAS 6-m telescope (Afanasiev et
al., 2004). The MPFS takes simultaneous spectra from 256 spatial
elements that collects an area of $16\times16''$ on the sky (see
full description at http://www.sao.ru/hq/lsfvo/devices.html). The
spectral interval was  $4500- 7100$\AA\, with resolution of 7-8
\AA\ (in 2002) and  $4350-5900$\AA\, with resolution of 4.5\AA\,
(in 2004).

\begin{figure}[]
\plottwo{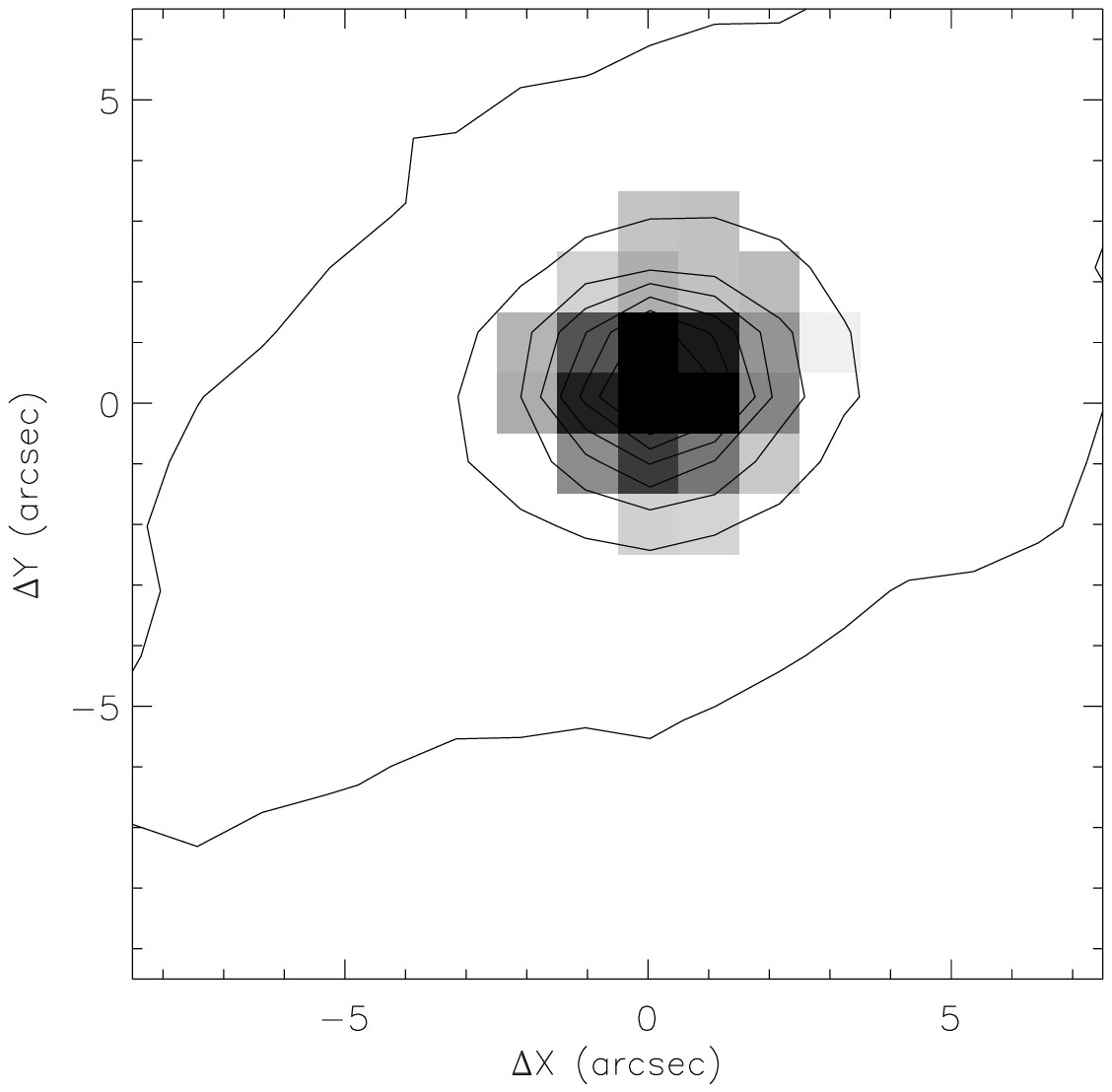}{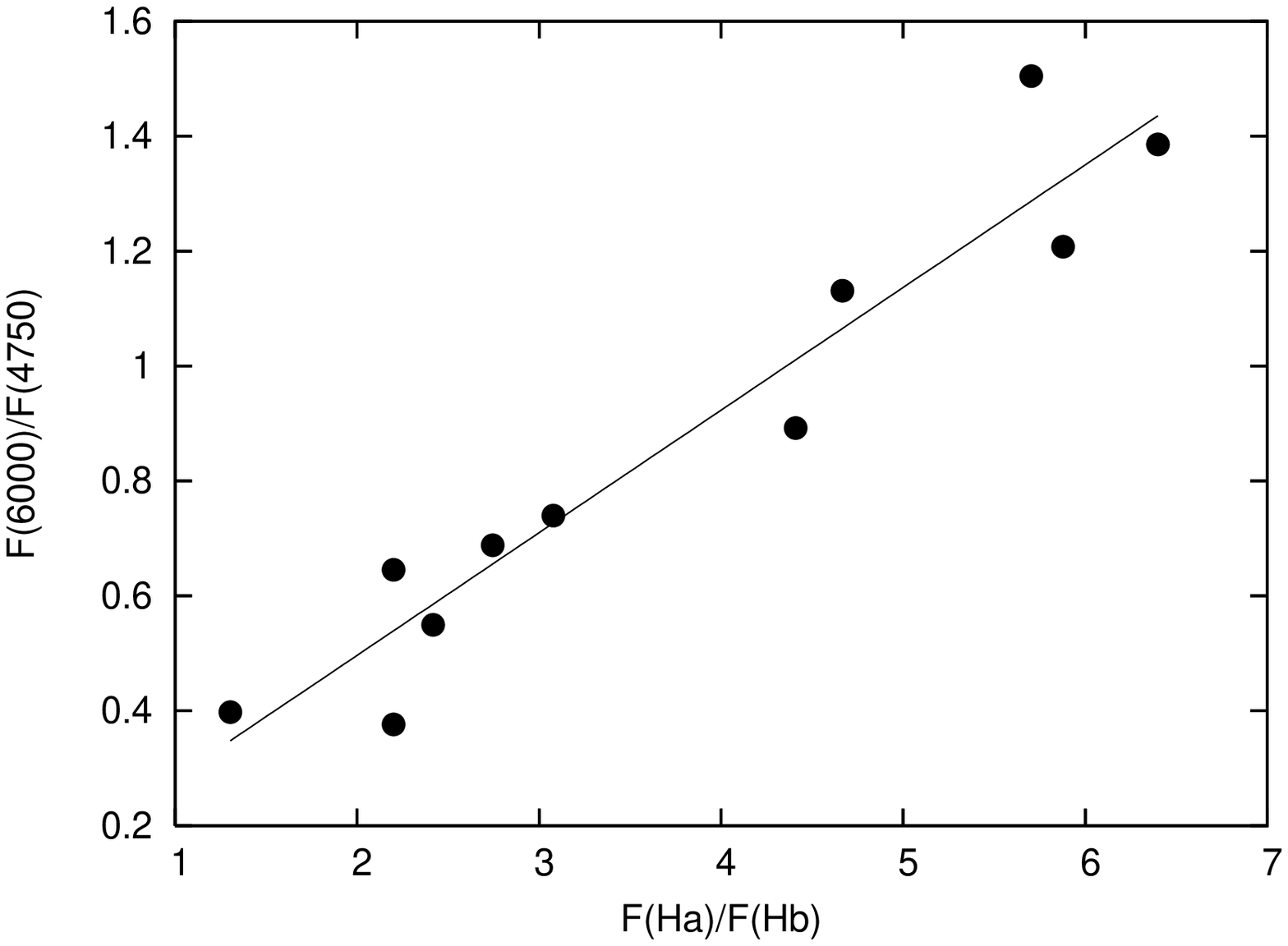}
\caption{The map of the Fe II line intensity and contours of the
continuum around 4900\AA\, across the Mrk 493 nucleus (left). The
ratio of the continuum vs. H$\alpha$/H$\beta$ ratio (right).}
\end{figure}

\section{Results}

By analyzing of the 3D spectroscopical observations:

i) we found that there is a strong Fe II emission in an extensive
region around Sy 1 nucleus (around $4''\times4''$ around the
nucleus, see Fig. 1, left).

ii) we detect
a strong difference in the slope of the optical  continuum emission
intensity across the nucleus part.

iii) the continuum ratio of the the red (6000 \AA ) and blue part (4750 \AA )
is in the correlation with the intensity ratio of H$\alpha$/H$\beta$
across the nucleus part (see Fig. 1, right).

We fitted the Fe II and H$\beta$ lines in nucleus region in order
to compare  the kinematic parameters of the Balmer and Fe II line
emitting regions.   We found that the broad H$\beta$ can be
decomposed into two components (broad -- W$_B$ and intermediate --
W$_I$) indicating the random velocity of emitting gas
around W$_B=$2400 km/s and W$_I=$600 km/s. On the other hand, the
widths of Fe II lines indicate the velocity gas in the Fe II
emission region around W$_{FeII}=$800 km/s, that is closer to the
intermediate region. This indicates that Fe II lines originate in
an intermediate line region as it was mentioned in Popovi\'c et
al. (2004)


\acknowledgements 
This work was supported by the Ministry
of Science and Environment Protection of Serbia through the project 146002
``Astrophysical Spectroscopy of Extragalactic Objects'' and by the Russian
Foundation for Basic Research (project nos. 06-02-16825).


\end{document}